# طراحی کنترل‌کننده پیش‌بین مدل پایدارساز برای سیستم‌های هایبرید مرکب منطقی دینامیکی: رویکرد تابع لیاپانف مبتنی بر نُرم بی‌نهایت

علیرضا علما[1]، کارشناس ارشد، مختار شاصادقی[2]، دانشیار، امین رمضانی[3]، استادیار

۱- دانشکده مهندسی برق و الکترونیک - دانشگاه صنعتی شیراز - شیراز - ایران – a.olama@sutech.ac.ir
۲- دانشکده مهندسی برق و الکترونیک - دانشگاه صنعتی شیراز - شیراز - ایران – shasadeghi@sutech.ac.ir
۳- دانشکده مهندسی برق و کامپیوتر – دانشگاه تربیت مدرس - تهران - ایران – ramezani@modares.ac.ir

**چکیده:** کنترل پیش‌بین سیستم‌های هایبرید با دو چالش اساسی تضمین پایداری حلقه-بسته و همچنین کاهش پیچیدگی محاسباتی روبه‌رو می‌باشد. در این مقاله، پایداری نمایی حلقه-بسته سیستم‌های هایبرید توصیف‌شده توسط مدل مرکب منطقی دینامیکی توسط کنترل پیش‌بین مدل تحلیل می‌شود. برای این منظور، استفاده از شرط نزولی‌بودن یک تابع لیاپانوف مبتنی بر نرم بی‌نهایت متغیرهای حالت سیستم، به‌جای تحمیل قید مساوی نهایی در مساله کنترل پیش‌بین مدل سیستم‌های مرکب منطقی دینامیکی پیشنهاد می‌شود. شرایط تضمین پایداری نمایی حلقه-بسته با استفاده از روش پیشنهادی دارای عملکرد بهتری هم از نظر پیاده‌سازی کنترل‌کننده و هم از نظر پیچیدگی محاسباتی است . علاوه‌بر این، با استفاده از این روش، شرایط پایداری نمایی حلقه-بسته نقطه تعادل به مقدار افق پیش‌بینی سیستم وابسته نمی‌باشد و همین امر می‌تواند یکی از مهمترین مزایای این روش در نظر گرفته شود. با استفاده از شرط نزولی‌بودن تابع لیاپانوف در فرمول‌بندی کنترل پیش‌بین مدل برای سیستم‌های مرکب منطقی دینامیکی، نسخه زیربهینه سیگنال کنترل با حجم محاسباتی بسیار کمتر به‌دست می‌آید. به‌منظور بررسی عملکرد روش پیشنهادشده، مساله پایدارسازی در سیستم تعلیق خودرو مورد مطالعه و شبیه‌سازی قرار می‌گیرد.

**واژه‌های کلیدی:** کنترل‌کننده پیش‌بین مدل، سیستم‌های هایبرید مرکب منطقی دینامیکی، پایداری، برنامه ریزی صحیح-مرکب

## Stable MPC Design for Hybrid Mixed Logical Dynamical Systems: $l\infty$-based Lyapunov Approach

A. Olama, MSc[1], M. Shasadeghi, Associate Professor[2], A. Ramezani, Assistant Professor[3]

1- Faculty of Electrical and Electronic Engineering, Shiraz University of Technology, Shiraz, Iran, Email: a.olama@sutech.ac.ir
2- Faculty of Electrical and Electronic Engineering, Shiraz University of Technology, Shiraz, Iran, Email: shasadeghi@sutech.ac.ir
3- Faculty of Electrical and Computer Engineering, Tarbiat Modares University, Tehran, Iran, Email: ramezani@modares.ac.ir

**Abstract:** There are two main challenges in control of hybrid systems which are to guarantee the closed-loop stability and reduce computational complexity. In this paper, we propose the exponential stability conditions of hybrid systems which are described in the Mixed Logical Dynamical (MLD) form in closed-loop with Model Predictive Control (MPC). To do this, it is proposed to use the decreasing condition of infinity norm based Lyapunov function instead of imposing the terminal equality constraint in the MPC formulation of MLD system. The exponential stability conditions have a better performance from both implementation and computational points of view. In addition, the exponential stability conditions of the equilibrium point of the MLD system do not depend on the prediction horizon of MPC problem which is the main advantage of the proposed method. On the other hand, by using the decreasing condition of the Lyapunov function in the MPC setup, the suboptimal version of the control signal with reduced complexity is obtained. In order to show the capabilities of the proposed method, the stabilization problem of the car suspension system is studied.

**Keywords:** Model predictive control, Mixed logical dynamical system, Mixed integer programming

## 1- مقدمه

سیستم‌های دینامیکی هایبرید [1]، سیستم‌های دینامیکی هستند که از ترکیب سیستم‌های دینامیکی پیوسته-حالت و سیستم‌های دینامیکی رویداد گسسته[2] تشکیل شده‌اند. این کلاس خاص از سیستم‌ها، در دهه‌های اخیر، به‌دلیل کاربردهای فراوان در صنایع کنترل، بسیار مورد توجه واقع شده‌اند. برای مثال، در یک مدار الکتریکی مبتنی بر عملیات کلیدزنی، ولتاژ و جریان با توجه به قوانین الکتریکی حاکم بر مدار، به‌صورت پیوسته نسبت به زمان و همچنین به‌صورت ناپیوسته نسبت به موقعیت کلیدهای موجود در مدار تغییر می‌کنند. به‌طور کلی، بسیاری از کاربردهای دنیای حقیقی و همچنین بسیاری از سیستم‌ها، ذاتاً دارای ویژگی‌های یک سیستم هایبرید می‌باشند.

با توجه به این موضوع که سیستم‌های هایبرید از دو زیرسیستم زمان-محور و رویداد-محور تشکیل شده‌اند، مدل فضای حالت کلاسیک، دارای توانایی لازم برای مدل‌سازی این سیستم‌ها نمی‌باشد. بنابراین، توسعه روش‌های مدل‌سازی جدید به‌منظور تحلیل و طراحی این سیستم‌ها امری ضروری به نظر می‌رسد. به‌همین دلیل، تاکنون رویکردهای مختلفی برای مدل‌سازی سیستم‌های هایبرید در مقالات ارائه شده‌اند که می‌توان از اتوماتای هایبرید [3-5]، سیستم‌های تکه‌ای مستوی[2] [6]، سیستم‌های مرکب منطقی دینامیکی[3] [7]، سیستم‌های MMPS[4][8]، سیستم‌های مکمل [9] و سیستم‌های مکمل تعمیم‌یافته به‌عنوان مهمترین کلاس‌های مدل‌سازی سیستم‌های هایبرید نام برد. هرچند که عناصر تشکیل‌دهنده این مدل‌ها یکسان می‌باشند، با این وجود، این روش‌های مدل‌سازی، از دیدگاه پیچیدگی تحلیل و طراحی، دارای نقاط قوت و ضعف منحصر به فرد خود می‌باشند. در مرجع [10] به خوبی نشان داده شده است که تحت شرایطی این کلاس‌های مدل‌سازی معادل خواهند بود. مهمترین مزیت معادل‌سازی این روش‌ها استفاده از مزایای مدل‌های مختلف می‌باشد.

در میان تمامی این روش‌های مدل‌سازی، سیستم‌های MLD دارای ویژگی‌های بسیار مناسبی مانند فرمول‌بندی مناسب جهت طراحی کنترل‌کننده بهینه، تشخیص و شناسایی خطا، تحلیل کنترل‌پذیری و رویت‌پذیری می‌باشند. علاوه بر موارد ذکرشده، با استفاده از مدل MLD می‌توان دسته بسیار وسیعی از سیستم‌های هایبرید را نیز مدل‌سازی نمود. به‌طور خلاصه در این روش، سیستم هایبرید توسط ترکیب متغیرهای حقیقی و باینری تشکیل می‌شود. ارتباط این متغیرها نیز به‌صورت قیدهایی به سیستم اضافه می‌شود. علاوه بر سیستم‌های هایبرید، با استفاده از مدل MLD می‌توان دسته وسیعی از سیستم‌ها، مانند سیستم‌های خطی مقید، ماشین‌های حالت محدود، و سیستم‌های دارای متغیرهای باینری را مدل‌سازی نمود.

از آنجا که سیستم MLD یک سیستم غیرخطی مقید و دارای دینامیک ناپیوسته می‌باشد، طراحی کنترل‌کننده برای چنین سیستمی امر بسیار چالش‌برانگیزی می‌باشد. در میان انواع روش‌های طراحی کنترل‌کننده، کنترل‌کننده پیش‌بین مدل[5] به‌دلیل دارابودن توانایی‌های منحصر به فرد در حل مسائل کنترلی پیچیده، به‌طور گسترده در صنایع مختلف مورد استفاده قرار می‌گیرد. به‌عنوان نمونه، در مرجع [11] این استراتژی کنترلی به‌منظور کنترل سیستم‌های عملیات از راه دور مورد استفاده قرار گرفته است. در مرجع [12]، بر اساس برنامه‌ریزی چند پارامتری، کنترل‌کننده MPC به‌صورت خارج از خط و با حجم محاسباتی کم برای حل مساله ردیابی مورد استفاده قرار گرفته است. به‌طور خلاصه، MPC یک استراتژی کنترلی است که در آن، سیگنال کنترل در زمان حال، توسط حل بر-خط (در هر زمان نمونه‌برداری) یک مسئله کنترل بهینه حلقه-باز افق محدود به‌دست می‌آید. حل این مسئله که در آن از متغیرهای حالت سیستم در زمان حال به‌عنوان شرط اولیه مسئله استفاده می‌شود، به تولید یک دنباله کنترلی منجر می‌شود که بر اساس استراتژی افق پس‌رفتی[6]، فقط اولین عنصر این دنباله (سیگنال کنترل در زمان حال) به سیستم اعمال می‌شود. به‌عبارت دقیق‌تر، در MPC، با استفاده از مدل دینامیکی سیستم تحت کنترل، رفتار آینده بردار خروجی سیستم پیش‌بینی می‌شود و بر اساس این پیش‌بینی، در هر لحظه از زمان، کنترل‌کننده با استفاده از حل یک مسئله بهینه‌سازی ریاضی که منجر به تولید سیگنال‌های کنترل در لحظات آتی می‌شود، بهترین فرمان کنترلی را جهت مینیمم‌کردن یک معیار کارایی در حضور قیود فیزیکی تحمیل‌شده به سیستم، اتخاذ می‌کند و این روند در هر لحظه از زمان تکرار می‌شود [13, 14].

در مرجع [7]، این استراتژی کنترلی برای سیستم‌های MLD تعمیم داده شده است. مهمترین وجه تمایز بین کنترل پیش‌بین هایبرید[7] و غیرهایبرید در این است که در H-MPC، دنباله کنترلی از حل یک مسئله برنامه‌ریزی صحیح-مرکب[8] [15, 16] در هر لحظه از زمان به‌دست می‌آید که در حالت کلی، یک مسئله NP-Hard [17] می‌باشد. در ادبیات پیچیدگی محاسباتی، NP-Hard به مسائلی اطلاق می‌شود که دست کم به اندازه دشوارترین مساله در NP دشوار می‌باشند. به همین دلیل، از نقطه نظر پیچیدگی محاسباتی، محاسبه دنباله کنترلی در H-MPC امر بسیار دشوارتر از سایر کلاس‌های MPC می‌باشد. تاکنون مقالات بسیار متنوعی از کاربردهای سیستم‌های MLD به همراه H-MPC معرفی شده است که می‌توان به کنترل سیستم‌های مرتبط با خودرو [18]، کنترل تحت شبکه [19] و کنترل شبکه‌های قدرت [22-20] اشاره نمود.

به‌طور کلی، H-MPC با دو چالش اساسی روبه‌رو می‌باشد که عبارتند از (1) تضمین پایداری حلقه-بسته و (2) کاهش پیچیدگی محاسباتی. در MPC غیرهایبرید، روش‌های مختلفی برای تضمین پایداری حلقه-بسته ارائه شده است. در یک دسته بسیار کاربردی از این روش‌ها، که به کنترل پیش‌بین با دو مود عملکردی شناخته می‌شود، یک کنترل‌کننده محلی وجود دارد که یک مجموعه پایای مثبت [23] حول نقطه تعادل سیستم ایجاد نموده و MPC سعی می‌کند بردار

حالت را به درون این مجموعه هدایت کند [۱۳]. در درون این مجموعه، کنترل‌کننده محلی وظیفه پایدارسازی سیستم را به عهده خواهد داشت. در مراجع [۲۴-۲۶]، این روش به‌منظور پایدارسازی سیستم‌های هایبرید مبتنی بر مدل PWA تعمیم داده شده است. اولین روش تضمین پایداری حلقه-بسته برای سیستم‌های MLD در مرجع [۷] معرفی شده است. در این روش که نوع سخت‌گیرانه‌تری از کنترل پیش‌بین با دو مد عملکرد است، مجموعه نهایی حول نقطه تعادل سیستم دقیقا نقطه تعادل در نظر گرفته می‌شود و همچنین کنترل‌کننده محلی نیز برابر صفر می‌باشد. دلیل این امر این است که به‌دلیل وجود متغیرهای باینری در معادلات دینامیکی سیستم، محاسبه کنترل‌کننده محلی و همچنین مجموعه پایا حول مبدا کار بسیار مشکلی می‌باشد. در این رویکرد، که به آن کنترل پیش‌بین با قید مساوی نهایی[۹] گفته می‌شود، بردار حالت سیستم می‌بایستی در انتهای افق پیش‌بینی به نقطه تعادل سیستم برسد. مهم‌ترین نقطه ضعف این روش در این است که این قید در بسیاری از موارد ارضاشدنی نیست. به‌عبارت دیگر، نقطه تعادل سیستم در چند گام زمانی قابل دسترسی نیست و این قید صرفا به‌ازای افق‌های پیش‌بینی بزرگ ارضاشدنی خواهد بود. علاوه‌بر این، به‌منظور برقراری این قید بدون افزایش افق پیش‌بینی، شرایط اولیه سیستم می‌بایستی به‌طور قابل‌توجهی به نقطه تعادل سیستم نزدیک باشند، از این‌رو، با توجه به NP-Hard بودن مسئله بهینه‌سازی، افزایش افق پیش‌بینی موجب افزایش ابعاد مساله و به تبع آن افزایش پیچیدگی محاسباتی مسئله می‌شود. از طرف دیگر، این قید در صورت ارضاشدن نیز تنها جاذب‌بودن نقطه تعادل را تضمین می‌کند و نمی‌توان با استفاده از آن پایداری لیاپانوف را اثبات نمود. از این جهت ارائه روش‌های تضمین‌کننده پایداری برای سیستم‌های MLD بدون استفاده از قید مساوی نهایی جهت کاهش پیچیدگی محاسباتی امر ضروری به نظر می‌رسد.

در این مقاله، به‌منظور تضمین پایداری نمایی حلقه-بسته و همچنین کاهش پیچیدگی محاسباتی مساله پایداری سیستم‌های MLD، از قید نزولی‌بودن تابع لیاپانوف به جای قید مساوی نهایی در مساله H-MPC استفاده شده است. برای این منظور، یک تابع لیاپانوف مبتنی بر نرم بی‌نهایت برای سیستم تعریف شده و H-MPC سعی می‌کند در هر گام زمانی، تابع لیاپانوف را کوچکتر نماید. در ادامه نشان داده خواهد شد که این قید به افق پیش‌بین مسئله وابسته نبوده و می‌توان با استفاده از آن، پایداری نمایی سیستم حلقه-بسته را به‌ازای افق پیش‌بینی بسیار کوچک (حتی برابر با یک) تضمین کرد. با توجه به رابطه مستقیم پیچیدگی محاسباتی با افق پیش‌بین مسئله H-MPC می‌توان با استفاده از روش پیشنهادی، پیچیدگی محاسباتی مسئله را نیز به‌طور چشم‌گیری کاهش داد. علاوه‌بر این موارد، با استفاده از ساختار قید نزول تابع لیاپانوف، همان‌گونه که نشان داده خواهد شد،

می‌توان نوع زیربهینه این کنترل‌کننده با پیچیدگی محاسباتی بسیار کمتر را نیز ارائه نمود.

## ۲- مقدمات

### ۲-۱- سیستم‌های مرکب منطقی-دینامیکی

ایده اصلی مدل‌سازی به روش MLD براساس تبدیل گزاره‌های منطقی به نامساوی‌های صحیح-مرکب می‌باشد [۲۷]. این نامساوی‌ها عموماً متشکل از متغیرهای صحیح و حقیقی می‌باشند. با استفاده از این روش، سیستم هایبرید را می‌توان به‌صورت یک سیستم شبه-خطی به‌همراه قیود صحیح-مرکب بیان نمود. برای مثال، عبارت منطقی "یا" را می‌توان توسط نامساوی صحیح مرکب معادل به‌صورت زیر بیان نمود:

$$X_1 \vee X_2 \equiv \delta_1 + \delta_2 \geq 1 \tag{۱}$$

که در آن $\delta_i \in \{0,1\}$ متغیر باینری به اشتراک گذاشته‌شده با متغیر منطقی $X_i$ می‌باشد که در صورت صحیح بودن عبارت منطقی مقدار یک و در غیر این صورت مقدار صفر به خود خواهد گرفت. رابطه (۱) این مفهوم را بیان می‌کند که عبارت منطقی سمت چپ صحیح است اگر و فقط اگر نامساوی عدد صحیح سمت راست برقرار باشد. با استفاده از این تکنیک و همچنین معرفی متغیرهای کمکی پیوسته به‌منظور تبدیل حاصل‌ضرب عبارات باینری و حقیقی به نامساوی‌های صحیح-مرکب، مدل MLD به‌صورت زیر به‌دست می‌آید:

$$x(t+1) = Ax(t) + B_1 u(t) + B_2 \delta(t) + B_3 z(t) \tag{۲}$$

$$y(t) = Cx(t) + D_1 u(t) + D_2 \delta(t) + D_3 z(t) \tag{۳}$$

$$E_2 \delta(t) + E_3 z(t) \leq E_1 u(t) + E_4 x(t) + E_5 \tag{۴}$$

در این روابط $t$ زمان گسسته،

$x = \begin{pmatrix} x_c \\ x_l \end{pmatrix}$, $x_c \in \mathcal{R}^{n_c}$, $x_l \in \{0,1\}^{n_l}$, $n \triangleq n_c + n_l$

بردار حالت سیستم شامل متغیرهای باینری و حقیقی،

$y = \begin{pmatrix} y_c \\ y_l \end{pmatrix}$, $y_c \in \mathcal{R}^{p_c}$, $y_l \in \{0,1\}^{p_l}$, $p \triangleq p_c + p_l$

بردار خروجی سیستم و

$u = \begin{pmatrix} u_c \\ u_l \end{pmatrix}$, $u_c \in \mathcal{R}^{m_c}$, $u_l \in \{0,1\}^{m_l}$, $m \triangleq m_c + m_l$

بردار ورودی سیستم شامل متغیرهای ورودی پیوسته و باینری می‌باشد. بردار $\delta \in \{0,1\}^{r_l}$ و $z \in \mathcal{R}^{r_c}$ به‌ترتیب، بردارهای متغیرهای کمکی باینری هستند که به‌عنوان ورودی‌های کنترلی سیستم MLD مورد استفاده قرار می‌گیرند. بنابراین روابط (۲) و (۳) به‌ترتیب، معادلات حالت و خروجی سیستم و رابطه (۴) قیدهای به‌دست آمده در طی فرایند مدل‌سازی را بیان می‌کند. ذکر این نکته لازم است که عموما برای یک سیستم هایبرید، مدل‌های MLD متفاوتی می‌توان ارائه نمود. دلیل این امر این است که معمولا نظیرنمودن متغیرهای باینری و نیز انتخاب متغیرهای کمکی دارای انتخاب‌های متفاوتی است

و هر یک از این انتخاب‌ها منجر به مدل MLD جداگانه‌ای می‌شود. به‌منظور جزئیات بیشتر به مرجع [7] رجوع شود.

## 2-2- تعاریف اساسی سیستم‌های MLD

تعاریف اساسی پایداری سیستم‌های MLD به‌صورت زیر می‌باشند[7]:

**تعریف1:** بردار $x_e \in \mathcal{R}^{nc} \times \{0,1\}^{nl}$ یک حالت تعادل[10] برای سیستم MLD توصیف‌شده توسط روابط (2) تا (4) و ورودی $u_e \in \mathcal{R}^{mc} \times \{0,1\}^{ml}$ نامیده می‌شود، اگر $[x_e^T \; u_e^T] \in \mathcal{F}$ و $x(t, x_e) = x_e \; \forall \; t \geq t_0$. در این صورت جفت $(x_e, u_e)$، جفت تعادل نامیده می‌شود. ناحیه $\mathcal{F}$، یک چندوجهی می‌باشد که سیستم MLD در آن تعریف می‌شود.

**تعریف 2:** [7] برای جفت تعادل $(x_e, u_e)$، $x_e$ پایدار است اگر برای $t_0 \in Z$ و به‌ازای هر $\epsilon \geq 0$، $\delta(\epsilon, t_0)$ به‌نحوی وجود داشته باشد که:

$$\|x_0 - x_e\| \leq \delta \Rightarrow \|x(x_0, u_e) - x_e\| \leq \epsilon, \quad \forall t \geq t_0 \tag{5}$$

**تعریف 3:** [7] برای جفت تعادل $(x_e, u_e)$، $x_e$ پایدار مجانبی است اگر $x_e$ پایدار باشد و $r > 0$ به‌نحوی وجود داشته باشد که به‌ازای هر $x_0 \in B(x_e, r)$ و $\epsilon > 0$، $T(\epsilon, t_0)$ تعریف شده و رابطه زیر برقرار باشد:

$$\|x(t) - x_e\| \leq \epsilon, \quad \forall t \geq T \tag{6}$$

**تعریف 4:** [7] برای جفت تعادل $(x_e, u_e)$، $x_e$ پایدار نمایی است اگر $x_e$ پایدار مجانبی باشد و $\delta > 0$، $\alpha > 0$ و $1 \leq \beta \leq 0$ به‌نحوی وجود داشته باشند که به‌ازای هر $x_0 \in B(x_e, \delta)$ رابطه زیر برقرار باشد:

$$\|x(t) - x_e\| \leq \alpha \beta^{t-t_0} \|x(t_0) - x_e\| \tag{7}$$

## 2-3- کنترل پیش‌بین مدل هایبرید

مساله کنترل پیش‌بین مدل برای سیستم‌های MLD در حالت کلی به‌صورت زیر تعریف می‌شود [7]:

$$\min_{U_t} \; J(U_t, x(t)) \tag{8}$$

$$x_{t+k+1|t} = Ax_{t+k|t} + B_1 u_{t+k|t} + B_2 \delta_{t+k|t} + B_3 z_{t+k|t} \tag{9}$$

$$y_{t+k|t} = Cx_{t+k|t} + D_1 u_{t+k|t} + D_2 \delta_{t+k|t} + D_3 z_{t+k|t} \tag{10}$$

$$E_2 \delta_{t+k|t} + E_3 z_{t+k|t} \leq E_1 u_{t+k|t} + E_4 x_{t+k|t} + E_5 \tag{11}$$

$$x_{t+N|t} = x_e \tag{12}$$

$$U_t = \{u_{t|t}, \ldots, u_{t+N-1|t}, \delta_{t|t}, \ldots, \delta_{t+N-1|t}, z_{t|t}, \ldots, z_{t+N-1|t}\} \tag{13}$$

که در رابطه (8)، $J(U_t, x(t))$ تابع هزینه مسئله H-MPC می‌باشد که به‌صورت کلی زیر تعریف می‌گردد:

$$J(U_t, x(t)) \triangleq \sum_{k=0}^{N-1} \|u_{t+k|t} - u_e\|_{Q_1}^2 + \|\delta_{t+k|t} - \delta_e\|_{Q_2}^2$$
$$+ \|z_{t+k|t} - z_e\|_{Q_3}^2 + \|x_{t+k|t} - x_e\|_{Q_4}^2 \tag{14}$$
$$+ \|y_{t+k|t} - y_e\|_{Q_5}^2$$

که در آن، $N$ افق پیش‌بینی، $U_t$ بردار متغیرهای بهینه‌سازی شامل ورودی‌کنترل، متغیر کمکی باینری و پیوسته، $y_e$، $x_e$، $u_e$، $z_e$ و $\delta_e$ مقادیر حالت ماندگار برای متغیر خروجی، حالت، ورودی و متغیرهای کمکی باینری و پیوسته و ماتریس‌های مثبت معین $Q_i, i = 1, \ldots 5$ ماتریس‌های وزنی با ابعاد مناسب می‌باشند. در این فرمول‌بندی، قید مساوی نهایی (12) به‌منظور تضمین پایداری حلقه-بسته مورد استفاده قرار گرفته است. پس از حل مساله بهینه‌سازی فوق در هر لحظه، طبق استراتژی RHC، اولین عنصر از بردار بهینه‌سازی به سیستم MLD اعمال می‌شود و برای لحظات بعد، افق پیش‌بینی در طول محور زمان شیفت داده شده و روند بهینه‌سازی مجدد تکرار می‌گردد.

## 2-4- بیان مساله

مساله H-MPC تعریف‌شده توسط روابط (5) تا (10) را با هدف پایداری در نقطه تعادل بدون استفاده از قید مساوی نهایی را به‌صورت زیر در نظر بگیرید:

$$\min_{U_t} \; J(U_t, x(t)) \tag{15}$$

$$x_{t+k+1|t} = Ax_{t+k|t} + B_1 u_{t+k|t} + B_2 \delta_{t+k|t} + B_3 z_{t+k|t} \tag{16}$$

$$y_{t+k|t} = Cx_{t+k|t} + D_1 u_{t+k|t} + D_2 \delta_{t+k|t} + D_3 z_{t+k|t} \tag{17}$$

$$E_2 \delta_{t+k|t} + E_3 z_{t+k|t} \leq E_1 u_{t+k|t} + E_4 x_{t+k|t} + E_5 \tag{18}$$

$$U_t = \{u_{t|t}, \ldots, u_{t+N-1|t}, \delta_{t|t}, \ldots, \delta_{t+N-1|t}, z_{t|t}, \ldots, z_{t+N-1|t}\} \tag{19}$$

هدف اصلی کنترل در این مساله استخراج دنباله کنترلی $u(t)$ و همچنین متغیرهای کمکی قابل قبول $\delta(t)$ و $z(t)$ به‌نحوی است که علاوه بر رعایت محدودیت‌های موجود در مسئله، پایداری نمایی سیستم MLD معرفی‌شده توسط روابط (2) تا (4) را در نقطه تعادل تضمین و همچنین تابع هزینه تعریف‌شده در رابطه (14) را کمینه نماید. در بخش بعد روش پیشنهادی به‌منظور حل مساله فوق به‌تفضیل بیان خواهد شد.

## 3- حل مساله پایدارسازی سیستم MLD با استفاده از H-MPC

در این بخش، با معرفی یک تابع لیاپانوف مبتنی بر نرم بی‌نهایت نشان داده خواهد شد که با اعمال قید نزولی‌بودن این تابع می‌توان بدون وابستگی به افق پیش‌بین و محاسبات مربوط به هزینه نهایی و مجموعه نهایی، پایداری نمایی حلقه بسته سیستم MLD را در نقطه تعادل تضمین نمود. بدیهی است که با اعمال تغییرات بسیار کمی در فرآیند آتی، می‌توان مساله ردیابی را نیز مورد بررسی قرار داد.

به‌منظور تضمین پایداری نمایی حلقه-بسته سیستم MLD تابع زیر به‌عنوان تابع لیاپانوف برای سیستم MLD در نظر گرفته می‌شود:

$$V(x(t)) = \|Yx(t)\|_\infty \qquad (20)$$

که در آن، $Y \in \mathcal{R}^{c \times n}, Rank(Y) = n$ می‌باشد. این تابع لیاپانوف برای سیستم‌های LTI در مرجع [28] و همچنین برای سیستم‌های هایبرید مبتنی بر مدل PWA در مراجع [29, 30] معرفی و استفاده شده است. در ادامه به حل مساله پایدارسازی در H-MPC با استفاده از این تابع لیاپانوف پرداخته خواهد شد.

**فرض 1:** $(x_e, u_e) = (0,0)$ حالت تعادل سیستم MLD می‌باشد.

با در نظر گرفتن فرض 1 و همچنین اعمال قید نزولی‌بودن تابع لیاپانوف موجود در رابطه (20)، مساله H-MPC را می‌توان به‌صورت زیر بیان نمود:

$$\min_{U_t} J(U_t, x(t)) \qquad (21)$$

$$x_{t+k+1|t} = Ax_{t+k|t} + B_1 u_{t+k|t} + B_2 \delta_{t+k|t} + B_3 z_{t+k|t} \qquad (22)$$

$$y_{t+k|t} = Cx_{t+k|t} + D_1 u_{t+k|t} + D_2 \delta_{t+k|t} + D_3 z_{t+k|t} \qquad (23)$$

$$E_2 \delta_{t+k|t} + E_3 z_{t+k|t} \leq E_1 u_{t+k|t} + E_4 x_{t+k|t} + E_5 \qquad (24)$$

$$\|Yx_{t+1|t}\|_\infty \leq \|Yx_{t|t}\|_\infty - \gamma \|x_{t|t}\|_\infty \qquad (25)$$

$$U_t = \{u_{t|t}, \ldots, u_{t+N-1|t}, \delta_{t|t}, \ldots, \delta_{t+N-1|t}, z_{t|t}, \ldots, z_{t+N-1|t}\} \qquad (26)$$

که در آن $\gamma > 0$ مقدار ثابت می‌باشد.

**تبصره 1:** نامساوی بیان‌شده در رابطه (25) قید نزولی‌بودن تابع لیاپانوف (20) را بیان می‌کند بدین مفهوم که تابع لیاپانوف می‌بایست در هر گام از گام قبل کوچکتر باشد. با استفاده از این قید، بردار حالت‌های سیستم در هر گام زمانی به نقطه تعادل نزدیکتر می‌شود و بر خلاف قید مساوی نهایی نیازی نیست که حالت‌های سیستم در انتهای افق پیش‌بینی به نقطه تعادل برسند. با توجه به اینکه در رابطه (25) مقادیر حالت‌های سیستم MLD فقط در زمان حال مورد استفاده قرار می‌گیرند این قید به افق پیش‌بین وابسته نمی‌باشد.

قبل از بیان قضیه پایدارسازی قید نزول تابع لیاپانوف (25)، فرض زیر در نظر گرفته می‌شود:

**فرض 2:** مجموعه $\Gamma \in \mathcal{F}$ وجود دارد به‌نحوی که به ازای $x(0) \in \Gamma$ مساله بهینه‌سازی تعریف‌شده توسط روابط (18) تا (23)، برای تمامی زمان‌های آتی دارای جواب باشد.

**تبصره 2:** به مجموعه $\Gamma$ مجموعه پایای انقباضی گفته می‌شود که روش محاسبه آن برای سیستم‌های LTI در مرجع [31] معرفی شده است. با توجه به اینکه محاسبه این مجموعه به متغیرهای باینری سیستم MLD وابسته نمی‌باشد می‌توان روش موجود در مرجع فوق را برای سیستم‌های MLD تعمیم داد. بنابراین فرض 2 محدودکننده‌ای نمی‌باشد.

**قضیه 1:** با در نظرگرفتن فرض‌های 1 و 2، سیستم MLD تعریف‌شده توسط روابط (2) تا (4) به‌صورت حلقه-بسته با قانون کنترل به‌دست‌آمده توسط مساله بهینه‌سازی تعریف‌شده در روابط (21) تا (26) و قاعده RHC، در درون مجموعه $\Gamma$، پایدار نمایی خواهد بود اگر و فقط اگر:

$$\|Y\|_\infty \geq \gamma \qquad (27)$$

$$\|Y\|_\infty \leq 1 + \gamma \qquad (28)$$

**اثبات:** با استفاده از قید موجود در رابطه (25) و فرض 2 و همچنین با استفاده از خاصیت زیرضربی در نرم بی‌نهایت می‌توان نوشت:

$$\|Yx_{t+1|t}\|_\infty \leq \|Y\|_\infty \|x(t)\|_\infty - \gamma \|x(t)\|_\infty \qquad (29)$$

$$\|Yx_{t+1|t}\|_\infty \leq (\|Y\|_\infty - \gamma) \|x(t)\|_\infty \qquad (30)$$

$$\|Yx_{t+1|t}\|_\infty \leq \theta \|x(t)\|_\infty \qquad (31)$$

با فرض $x(0) \in B(0, \lambda)$، $0 \leq \theta \leq 1$ خواهیم داشت:

$$\|Yx(t)\|_\infty \leq \theta \|x(t-1)\|_\infty \leq \theta^2 \|x(t-2)\|_\infty \leq \ldots \leq \theta^t \|x(0)\|_\infty \qquad (32)$$

رابطه (32) شرط برقراری پایداری نمایی را طبق تعریف 4 بیان می‌کند. به‌منظور برقراری قید $0 \leq \theta \leq 1$، ماتریس $Y$ باید دارای شرایط زیر نیز باشد:

$$\theta \geq 0 \rightarrow \|Y\|_\infty \geq \gamma \qquad (33)$$

$$\theta \leq 1 \rightarrow \|Y\|_\infty \leq 1 + \gamma \qquad (34)$$

بنابراین با توجه به اینکه قید پایدارسازی (25) در هرگام از مساله کنترل پیش‌بین، برای بردار حالت سیستم در دو گام متوالی $t$ و $t+1$ ارضاشدنی است، مبدا مختصات سیستم MLD با نرخ کاهش $\theta$ پایدار نمایی بوده و دامنه جذب آن مجموعه $\Gamma$ می‌باشد. ∎

به‌منظور محاسبه ماتریس $Y$ می‌توان از روش‌های بهینه‌سازی محدب [32] استفاده نمود.

**تبصره 3:** به‌عنوان یک نکته مهم می‌توان گفت در حالت کلی استفاده از قید نزول تابع لیاپانوف باعث وسیع‌ترشدن ناحیه جواب مساله بهینه‌سازی در H-MPC می‌شود. دلیل این امر آن است که برخلاف قید مساوی نهایی، در این فرمول‌بندی، H-MPC سعی می‌کند که در هر گام بردار حالت‌های سیستم را به نقطه تعادل نزدیکتر کند در صورتی که قید مساوی نهایی، قید بسیار سخت‌گیرانه‌تری نسبت به این قید می‌باشد. علاوه بر این، با توجه به اینکه ناحیه گسترده‌تری از شرایط اولیه سیستم منجر به برقراری قید نزولی‌بودن تابع لیاپانوف می‌گردند، ناحیه جذب سیستم با استفاده از این کنترل‌کننده نسبت به کنترل‌کننده مبتنی بر قید مساوی نهایی وسیع‌تر خواهد بود.

**تبصره 4:** مسائل بهینه‌سازی به‌دست‌آمده در فوق، این نکته را بیان می‌کنند که بردار حالت‌های سیستم، در هر گام زمانی، می‌بایست به نقطه تعادل سیستم نزدیک شود. بنابراین در صورتی که تابع هدف در این مسائل حذف شود، به‌دلیل وجود قید پایدارساز، همچنان بردار حالت‌های سیستم به سمت نقطه تعادل در حرکت خواهد بود. در نتیجه می‌توان در مسائل بهینه‌سازی فوق، با صرف‌نظر از بهترین جواب، صرفا به‌دنبال جواب قابل‌قبول بود و مساله بهینه‌سازی را به مساله امکان‌پذیری[11] به‌فرم زیر تبدیل نمود:

$$\underset{U_t}{Find}$$

$$x_{t+k+1|t} = Ax_{t+k|t} + B_1 u_{t+k|t} + B_2 \delta_{t+k|t} + B_3 z_{t+k|t} \quad (35)$$

$$y_{t+k|t} = Cx_{t+k|t} + D_1 u_{t+k|t} + D_2 \delta_{t+k|t} + D_3 z_{t+k|t} \quad (36)$$

$$E_2 \delta_{t+k|t} + E_3 z_{t+k|t} \leq E_1 u_{t+k|t} + E_4 x_{t+k|t} + E_5 \quad (37)$$

$$\|Yx_{t+1|t}\|_\infty \leq \|Yx_{t|t}\|_\infty - \gamma \|x_{t|t}\|_\infty \quad (38)$$

$$U_t = \{u_{t|t}, \ldots, u_{t+N-1|t}, \delta_{t|t}, \ldots, \delta_{t+N-1|t}, z_{t|t}, \ldots, z_{t+N-1|t}\} \quad (39)$$

جواب به‌دست‌آمده از مسائل فوق گرچه زیربهینه می‌باشد ولی به‌طور قابل ملاحظه‌ای دارای پیچیدگی محاسباتی کمتری نسبت به حالت بهینه است.

## 4- پیاده‌سازی کنترل‌کننده H-MPC

مساله بهینه‌سازی تعریف‌شده توسط روابط (21) تا (26) به‌طور مستقیم توسط الگوریتم‌های بهینه‌سازی ریاضی قابل پشتیبانی نبوده و می‌بایستی به فرم استاندارد مسائل بهینه‌سازی تبدیل شود. با استفاده از تکنیک معادل‌سازی قیودِ شامل نرم [32] و همچنین تعریف بردارها و ماتریس‌های مناسب می‌توان مساله بهینه‌سازی مذکور را به‌فرم استاندارد زیر بازنویسی کرد:

$$\underset{U_t}{\min} \quad J_t = \frac{1}{2} U_t^T H U_t + x(t)^T F U_t \quad (40)$$

$$\Phi U_t \leq \phi \quad (41)$$

که در آن، ماتریس $H$ و بردار $x(t)^T F$ به‌ترتیب ماتریس هسین و بردار گرادیان با ابعاد مناسب و همچنین بردار $U_t$ بردار متغیرهای تصمیم‌گیری مساله می‌باشد. رابطه (41) نیز بیانگر قیدهای مساله بهینه‌سازی است که شامل قیود عملکردی و همچنین قیود به‌دست‌آمده در خلال فرآیند مدل‌سازی می‌باشد. با توجه به وجود متغیرهای باینری در مساله بهینه‌سازی تعریف‌شده در روابط (40) و (41)، این مساله از نوع برنامه‌ریزی صحیح-مرکب درجه دو[12] [15, 16] می‌باشد.

این مسائل اکثرا در دسته مسائل NP-hard قرار می‌گیرند. در نتیجه در حالت کلی، مساله MIQP در زمان چندجمله‌ای[13] قابل حل نخواهند بود. ساده‌ترین راه برای حل چنین مسائلی بررسی تمامی جواب‌های موجود و انتخاب بهترین جواب می‌باشد که این کار در عمل به‌دلیل افزایش نمایی جواب‌های بالقوه مساله امکان‌پذیر نیست. تاکنون روش‌های مختلفی برای حل این مسائل ارائه شده‌اند که مهمترین آن‌ها عبارت‌اند از:

- روش‌های مبتنی بر برش صفحه
- روش‌های مبتنی بر تجزیه‌سازی
- روش تقریب بیرونی
- روش‌های شاخه و کران[14] [15، 16، 33].

در مرجع [34] به بررسی کارایی روش‌های فوق برای حل مساله MIQP پرداخته شده است که با توجه به نتایج موجود در این مرجع، از بین روش‌های ارائه‌شده، روش BnB دارای بهترین عملکرد می‌باشد. تمامی نرم‌افزارهای تجاری و غیر تجاری حل مسائل MIQP مانند CPLEX[35]، GUROBI [36] نیز بر اساس این روش کار می‌کنند.

## 5- کاربرد در حل مساله پایدارسازی در کنترل هایبرید سیستم تعلیق خودرو

دیاگرام بلوکی سیستم تعلیق دو درجه آزاد مجهز به یک نیروی قابل تنظیم با استفاده از روش مدل‌سازی موسوم به یک چهارم خودرو[15] [37] در شکل (1) نشان داده شده است [38]. با‌فرض خطی‌بودن فنرها، دمپر و همچنین با صرف‌نظر از اصطکاک چرخ‌ها و اغتشاش جاده، می‌توان رفتار دینامیکی سیستم تعلیق را توسط مدل فضای حالت زیر بیان کرد:

$$\dot{x} = Ax + B\bar{f} \quad (42)$$

$$A = \begin{pmatrix} 0 & 1 & 0 & 0 \\ -\omega_{us}^2 & -2\rho\zeta\omega_s & \rho\omega_s^2 & 2\rho\omega_s \\ 0 & -1 & 0 & 1 \\ 0 & 2\zeta\omega_s & -\omega_s^2 & -2\zeta\omega_s \end{pmatrix}, \quad B = \begin{pmatrix} 0 \\ \rho \\ 0 \\ -1 \end{pmatrix}$$

که در آن، $x = (x_1 \quad x_2 \quad x_3 \quad x_4)^T \in \mathcal{R}^4$ بردار حالت سیستم می‌باشد که در آن $x_1$ میزان انحراف چرخ‌ها[16] از نقطه تعادل، $x_2$ سرعت جرم $M_{us}$[17]، $x_3$ میزان انحراف تعلیق[18] از نقطه تعادل، $x_4$ سرعت جرم $M_s$[19] و $\bar{f}$ نیروی قابل تنظیم نرمال‌شده می‌باشد.

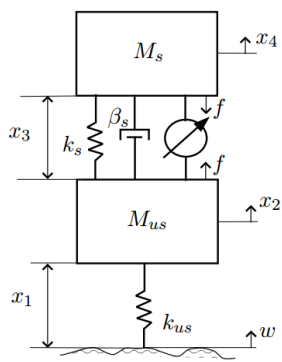

**شکل 1:** مدل یک‌چهارم خودرو سیستم تعلیق خودرو [38]

علاوه بر این، استفاده از نیروی قابل تنظیم $\bar{f}$ قیود زیر را به مسئله تحمیل می‌کند:

$$\bar{f}(x_4 - x_2) \geq 0 \tag{43}$$

$$|\bar{f}| \leq \sigma \tag{44}$$

$$\bar{f}(x_4 - x_2) \leq (2\zeta_{max}\omega_s)(x_4 - x_2)^2 \tag{45}$$

هدف کنترل در این سیستم، پایدارسازی حالت‌های سیستم در نقطه تعادل (مبدا) می‌باشد. از طرفی، همان‌گونه که از رابطه (43) و (45) مشخص است، این سیستم مقید به قیدهای نامحدب بوده و همین امر، کنترل مقید این سیستم را بسیار مشکل می‌کند. با استفاده از مدل‌سازی MLD می‌توان قیود نامحدب فوق را توسط متغیرهای باینری به قیود قطعه‌ای محدب تبدیل و با استفاده از این مدل طبق فرمول‌بندی پایدارساز ارائه‌شده در بخش قبل، الگوریتم H-MPC را اجرا کرد. برای این کار، ابتدا قید نامحدب رابطه (43) را در نظر بگیرید. با استفاده از متغیرهای باینری کمکی $\delta_1$ و $\delta_2$ می‌توان نوشت:

$$[\delta_1 = 1] \leftrightarrow [x_4 - x_2 \geq 0] \tag{46}$$

$$[\delta_2 = 1] \leftrightarrow [\bar{f} \geq 0] \tag{47}$$

$$[\delta_1 = 1] \rightarrow [\delta_2 = 1] \tag{48}$$

$$[\delta_1 = 0] \rightarrow [\delta_2 = 0] \tag{49}$$

به‌عبارت دقیق‌تر، توسط روابط (46) تا (49) می‌توان قید غیرخطی و نامحدب (43) را به صورت قیودی تکه‌ای محدب بازنویسی کرد. به‌عبارت دیگر، به‌عنوان نمونه، در صورتی که $x_4 - x_2 \geq 0$ باشد، $\delta_1 = 1$ خواهد بود و با توجه به رابطه (48) مقدار $\bar{f}$ می‌بایستی نامنفی باشد و این همان برقراری قید (43) می‌باشد. حالت‌های دیگر را نیز می‌توان به همین روش بررسی نمود. در حالت کلی، با استفاده از این روش می‌توان یک ناحیه نامحدب را به چند ناحیه محدب تبدیل نمود.

قید نامحدب موجود در نامساوی (45) را نیز می‌توان توسط متغیر کمکی پیوسته $F$ به‌صورت زیر بیان کرد:

$$\bar{f}(x_4 - x_2) \leq (2\zeta_{max}\omega_s)(x_4 - x_2)^2 \equiv F \geq 0 \tag{50}$$

$$F = \begin{cases} \bar{f} - (2\zeta_{max}\omega_s)(x_4 - x_2) & \text{if} \quad \delta_1 = 0 \\ -\bar{f} + (2\zeta_{max}\omega_s)(x_4 - x_2) & \text{if} \quad \text{otherwise} \end{cases} \tag{51}$$

که در اینجا، F متغیر کمکی پیوسته می‌باشد که در مدل MLD با نام $z(t)$ مشخص شده است. با توجه به اینکه معمولا فرآیند تبدیل گزاره‌های منطقی به قیود صحیح-مرکب کار بسیار زمان‌بری می‌باشد، در مرجع یک زبان برنامه نویسی به اسم HYSDEL [39] معرفی شده است که به وسیله آن می‌توان به‌راحتی مدل MLD سیستم هایبرید تحت مطالعه را استخراج کرد. جهت شبیه‌سازی این مساله، نرم‌افزار MATLAB به همراه حل‌کننده IBM CPLEX v12.6، به‌منظور حل مسئله MIQP، مورد استفاده واقع شده است. مقادیر عددی پارامترهای سیستم و همچنین پارامترهای استفاده‌شده در H-MPC در جدول (1) قابل مشاهده می‌باشند. در ادامه نتایج شبیه‌سازی سیستم تعلیق خودرو با شرط اولیه $x(0) = (0 \quad 0 \quad 0.1 \quad 0)^T$ بررسی خواهد شد.

در اینجا جهت مقایسه بهتر، سه حالت بهینه، زیربهینه و روش استفاده‌شده در مرجع [38][20] مورد بررسی قرار گرفته‌اند. مسیرهای حالت سیستم برای سه حالت مذکور به‌ترتیب در شکل‌های (2) تا (5) نمایش داده شده‌اند.

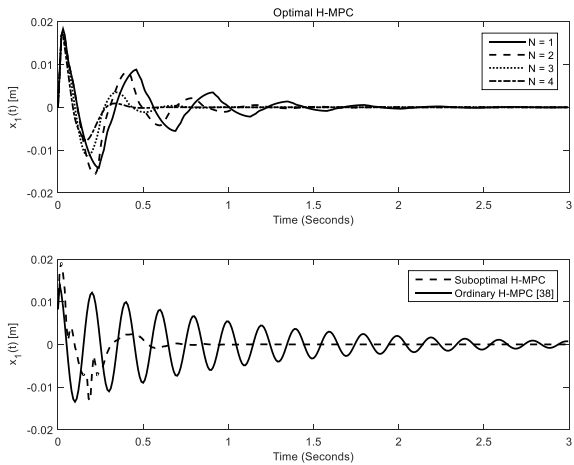

**شکل 2: مسیر حالت متغیر $x_1$**

در حالت بهینه، عملکرد کنترل‌کننده به‌ازای افق‌های پیش‌بین مختلف مورد بررسی قرار گرفته است. همانگونه که مشاهده می‌شود، سیستم حلقه‌بسته به‌ازای افق پیش‌بین $N = 1$ نیز به‌خوبی به نقطه تعادل همگرا شده است و بنابر آنچه در بخش‌های گذشته ذکر گردید، پایداری مستقل از افق پیش‌بینی قابل دسترسی خواهد بود. از طرف دیگر، با افزایش افق پیش‌بین می‌توان با حفظ پایداری حلقه بسته، پاسخ سیستم را بهبود داد.

**جدول 1: پارامترهای شبیه‌سازی سیستم تعلیق خودرو[38]**

| مقدار عددی | توضیحات | پارامتر |
|---|---|---|
| 9 ms | زمان نمونه‌برداری | $T_s$ |
| 1/5 Hz | فرکانس طبیعی جرم فنربندی‌شده $M_s$ | $\omega_s$ |
| 9 Hz | فرکانس طبیعی جرم فنربندی‌نشده $M_{us}$ | $\omega_{us}$ |
| 10 | نسبت جرم فنربندی‌شده به جرم فنربندی‌نشده | $\rho$ |
| 0 | ضریب میرایی | $\zeta$ |
| 2/25 | ضریب میرایی ماکزیمم | $\zeta_{max}$ |
| 5 | افق پیش‌بین | $N$ |
| 0/2 | حد بالای نیروی اشباع عملگر | $\sigma$ |

با توجه به اینکه، در کنترل‌کننده زیربهینه، از تابع هزینه صرف نظر می‌شود، افزایش افق پیش‌بین تاثیری در بهبود عملکرد حلقه-

بسته کنترل‌کننده زیربهینه ندارد و همین امر می‌تواند به‌عنوان یکی از محدودیت‌های این روش در نظر گرفته شود. میزان افق پیش‌بین برای این کنترل‌کننده برابر با ۵ در نظر گرفته شده است. در صـورتی کـه به‌منظور برقراری قید مساوی نهایی می‌بایست افق پیش‌بینی برابـر بـا ۱۳ باشد که در این صورت تعداد قیود صحیح-مرکب سیسـتم بسـیار افزایش یافته و به تبع آن پیچیدگی محاسباتی کنتـرل‌کننـده افـزایش خواهد یافت. علاوه‌بر موارد فوق، در مقایسه بـا روش اسـتفاده‌شـده در مرجع [۳۸]، روش پیشنهادی منجر به زمـان نشسـت بسـیار کمتـری گردیده است. از طرفی، با تنظیم مناسب افق پیش‌بین مساله، می‌توان از تحمیل حجم محاسبات جهت تضمین پایداری حلقـه بسـته تـا حـد مطلوبی جلوگیری نمود.

شکل (۶) نیروی قابل تنظیم اعمالی به سیستم (سـیگنال کنتـرل) را نمایش می‌دهد. همان‌گونه که مشاهده می‌شود سـیگنال کنتـرل در محدوده مجـاز تعریـف‌شـده در رابطـه (۴۴) قـرار دارد. کنتـرل‌کننـده زیربهینه در مقایسه با دو روش دیگـر دارای رفتـار نوسـانی در حالـت گذرای سیستم می‌باشد که دلیل این امر صرف‌نظر از تابع هزینه مساله می‌باشد. واضح است که نرخ همگرایـی هـر دو کنترل‌کننـده بهینـه و زیربهینه وضعیت بهتری نسبت به کنترل‌کننده طراحی‌شده بـر اسـاس روش‌های موجود دارد.

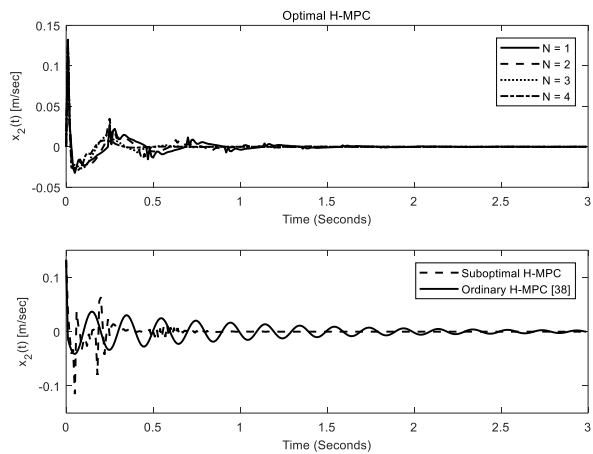

شکل ۳: مسیر حالت متغیر $x_2$

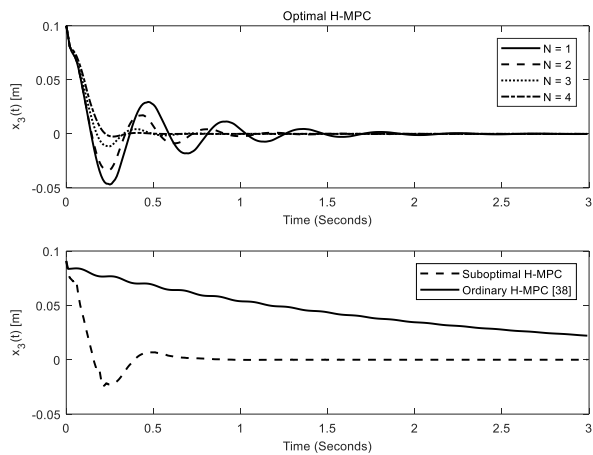

شکل ۴: مسیر حالت متغیر $x_3$

میزان زمان مصرفی به‌منظور حل مساله MIQP برای سه حالت بهینـه، زیربهینه و در حالت قید مساوی نهایی در هر زمان نمونـه بـرداری در شکل (۷) نمایش داده شده است.

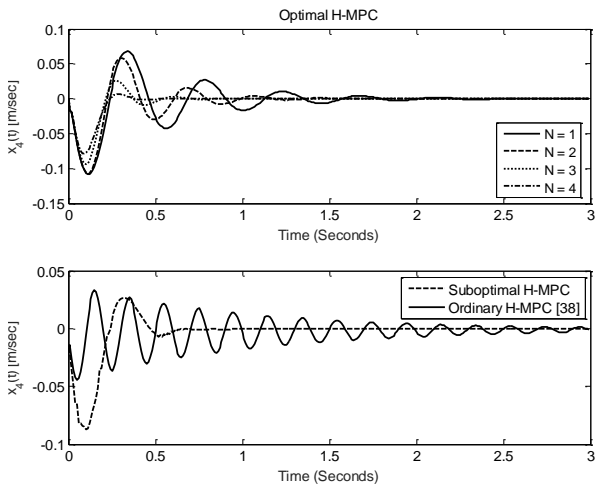

شکل ۵: مسیر حالت متغیر $x_4$

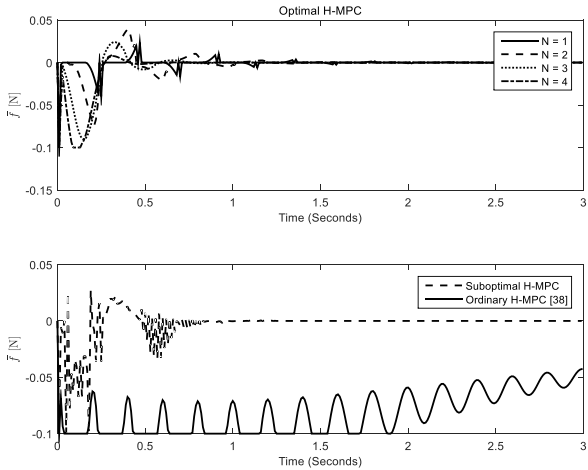

شکل ۶: سیگنال کنترل $\bar{f}$

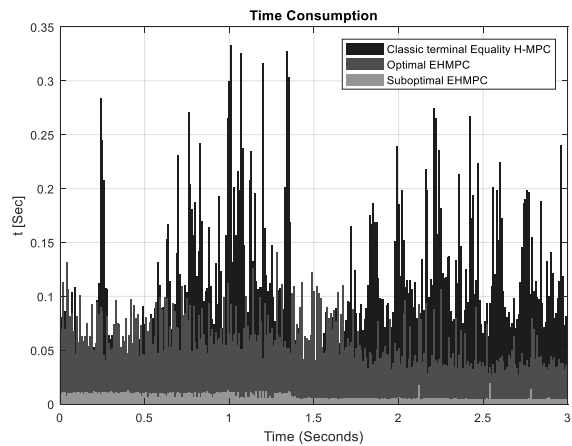

شکل ۷: نمودار زمان مصرفی در حل مساله MIQP

همان‌گونه که در شکل (۷) نشان داده شده است، کنتـرل‌کننـده زیربهینه به مراتب زمان کمتری نسبت به سایر کنترل‌کننـده‌هـا بـرای

حل مسأله MIQP نیاز دارد. در مقایسه با روش قید مساوی نهایی، هـر دو کنتـرل‌کننـده پیشــنهادی دارای پیچیــدگی محاســباتی کمتــری می‌باشند. با توجه به افزایش حجم مسأله در حالـت اسـتفاده از قیـد مساوی نهایی میزان زمان مصرفی به‌منظور حـل مسـأله MIQP بسـیار افزایش خواهد یافت.

## 6- نتیجه‌گیری

در این مقاله، حل مسئله پایدارسازی توسط کنترل پیش‌بـین بـرای سیستم‌های هایبرید مرکب منطقی-دینامیکی از ۲ جهت مـورد بهبـود واقع گردید. در درجه اول با استفاده از کنترل پیش‌بین هایبرید مبتنی بر تابع لیاپانوف می‌توان با محـدودیت کمتـری پایـداری حلقـه بسـته سیستم هایبرید را تضمین نمود. در مقایسه با کنترل پیش‌بین هایبرید مبتنی بر قید مساوی نهایی، قیـد نـزول تـابع لیاپـانوف از نظـر امکـان پیاده‌سازی و محاسبات کنترلی، سـهولت بیشـتری فـراهم مـی‌سـازد. دومین مزیت این روش در تبدیل مسأله بهینـه‌سـازی بـه یـک مسـأله امکان‌پذیری می‌باشـد. بـا توجـه بـه ایـن موضـوع کـه حـل مسـأله امکان‌پذیری از پیچیـدگی محاسـباتی کمتـری نسـبت بـه مسـأله بهینه‌سازی برخوردار است، بـا اسـتفاده از قیـد نـزول تـابع لیاپـانوف می‌توان با صرف‌نظر کردن از پاسخ بهینه و استفاده از پاسخ زیربهینـه، حجم محاسبات کنترل‌کننده را به‌طور چشم‌گیری کاهش داد. علاوه بر موارد فوق، بر خلاف قید مساوی نهایی که فقط جذبندگی نقطه تعادل را تضمین می‌نماید، بـا اسـتفاده از روش پیشــنهادی پایـداری نمـایی حلقه-بسته تضمین می‌گردد.

## مراجع

[1] M. S. Branicky, "Introduction to hybrid systems," *Handbook of networked and embedded control systems*, pp. 91-116, 2005.

[2] C. G. Cassandras and S. Lafortune, *Introduction to discrete event systems*. Springer Science & Business Media, 2009.

[3] T. A. Henzinger, P. W. Kopke, A. Puri and P. Varaiya, "What's decidable about hybrid automata?," in *Proceedings of the twenty-seventh annual ACM symposium on Theory of computing*, pp. 373-382, 1995.

[4] T. A. Henzinger, "The theory of hybrid automata," in *Verification of Digital and Hybrid Systems*: Springer, pp. 265-292, 2000.

[5] J. Lygeros, K. H. Johansson, S. N. Simic, J. Zhang, and S. S. Sastry, "Dynamical properties of hybrid automata," *IEEE Transactions on automatic control,* vol. 48, no. 1, pp. 2-17, 2003.

[6] F. J. Christophersen, "Piecewise affine systems," *Optimal Control of Constrained Piecewise Affine Systems,* pp. 39-42, 2007.

[7] A. Bemporad and M. Morari, "Control of systems integrating logic, dynamics, and constraints," *Automatica,* vol. 35, no. 3, pp. 407-427, 1999.

[8] B. De Schutter and T. Van den Boom, "Model predictive control for max-min-plus-scaling systems," in *American Control Conference, 2001. Proceedings of the 2001*, vol. 1, pp. 319-324: 2001.

[9] J. Schumacher, S. Weiland, and W. Heemels, "Linear complementarity systems," *SIAM journal on applied mathematics,* vol. 60, no. 4, pp. 1234-1269, 2000.

[10] W. P. Heemels, B. De Schutter, and A. Bemporad, "Equivalence of hybrid dynamical models," *Automatica,* vol. 37, no. 7, pp. 1085-1091, 2001.

[۱۱] سید وحید قوشخانه‌ای، علیرضا الفی، «طراحی کنترل پیش‌بین برای سیستم‌های عملیات از راه دور دوطرفه نامعین»، *مجله مهندسی برق دانشگاه تبریز*، دوره ۴۷، شماره ۲، ۱۳۹۶.

[۱۲] فرهاد بیات، صالح مبین، «طراحی کنترل پیش‌بین با هزینه محاسباتی کم: رویکرد برنامه ریزی پارامتری»، *مجله مهندسی برق دانشگاه تبریز*، دوره ۴۶، شماره ۴، ۱۳۹۵.

[13] J. B. Rawlings and D. Q. Mayne, *Model predictive control: Theory and design*. Nob Hill Pub., 2009.

[14] D. Q. Mayne, "Model predictive control: Recent developments and future promise," *Automatica,* vol. 50, no. 12, pp. 2967-2986, 2014.

[15] D. Bertsimas and R. Weismantel, *Optimization over integers*. Dynamic Ideas Belmont, 2005.

[16] L. A. Wolsey, *Integer programming*. Wiley, 1998.

[17] C. H. Papadimitriou, *Computational complexity*. John Wiley and Sons Ltd., 2003.

[18] A. Bemporad, N. Giorgetti, I. Kolmanovsky, and D. Hrovat, "A hybrid system approach to modeling and optimal control of DISC engines," in *Decision and Control, 2002, Proceedings of the 41st IEEE Conference on*, vol. 2, pp. 1582-1587, 2002.

[19] A. K. Sampathirao, P. Sopasakis, A. Bemporad, and P. Patrinos, "GPU-accelerated stochastic predictive control of drinking water networks," *IEEE Transactions on Control Systems Technology,* pp.551-562, 2017.

[20] A. G. Beccuti, T. Geyer, and M. Morari, "A hybrid system approach to power systems voltage control," in *Decision and Control, 2005 and 2005 European Control Conference. CDC-ECC'05. 44th IEEE Conference on*, pp. 6774-6779, 2005.

[21] M. Falahi, K. Butler-Purry, and M. Ehsani, "Dynamic reactive power control of islanded microgrids," *IEEE Transactions on Power Systems,* vol. 28, no. 4, pp. 3649-3657, 2013.

[22] W. Jing, C. H. Lai, S. H. W. Wong, and M. L. D. Wong, "Battery-supercapacitor hybrid energy storage system in standalone DC microgrids: areview," *IET Renewable Power Generation,* vol. 11, no. 4, pp. 461-469, 2016.

[23] F. Blanchini, "Set invariance in control," *Automatica,* vol. 35, no. 11, pp. 1747-1767, 1999.

[24] M. Lazar, W. Heemels, S. Weiland, and A. Bemporad, "Stabilizing model predictive control of hybrid systems," *IEEE Transactions on Automatic Control,* vol. 51, no. 11, pp. 1813-1818, 2006.

[25] M. Lazar and W. Heemels, "A new dual-mode hybrid MPC algorithm with a robust stability guarantee," *IFAC Proceedings Volumes,* vol. 39, no. 5, pp. 321-328, 2006.

[26] M. Lazar, "Model predictive control of hybrid systems: Stability and robustness," vol. 68, 2006.

[27] H. P. Williams, *Model building in mathematical programming*. John Wiley & Sons, 2013.

[28] A. Polanski, "On infinity norms as Lyapunov functions for linear systems," *IEEE Transactions on Automatic Control,* vol. 40, no. 7, pp. 1270-1274, 1995.

[29] M. Lazar and A. Jokić, "On infinity norms as Lyapunov functions for piecewise affine systems," in *Proceedings of the*

*13th ACM international conference on Hybrid systems: computation and control*, pp. 131-140, 2010.

[30] M. Lazar, W. Heemels, S. Weiland, A. Bemporad, and O. Pastravanu, "Infinity norms as Lyapunov functions for model predictive control of constrained PWA systems," *Lecture Notes in Computer Science,* vol. 3414, pp. 417-432, 2005.

[31] F. Blanchini, "Ultimate boundedness control for uncertain discrete-time systems via set-induced Lyapunov functions ",in *Decision and Control, 1991., Proceedings of the 30th IEEE Conference on*, pp. 1755-1760, 1991.

[32] S. Boyd and L. Vandenberghe, *Convex optimization*. Cambridge university press, 2004.

[33] J. Lee and S. Leyffer, *Mixed integer nonlinear programming*. Springer Science & Business Media, 2011.

[34] R. Fletcher and S. Leyffer, "Numerical experience with lower bounds for MIQP branch-and-bound," *SIAM Journal on Optimization,* vol. 8, no. 2, pp. 604-616, 1998.

[35] I. I. CPLEX, "V12. 1: User's Manual for CPLEX," *International Business Machines Corporation,* vol. 46, no. 53, p. 157, 2009.

[36] G. Optimization, "Inc.,"Gurobi optimizer reference manual," 2014," *URL: http://www. gurobi. com,* 2014.

[37] R. N. Jazar, "Quarter car," *Vehicle Dynamics: Theory and Application,* pp. 931-975, 2008.

[38] N. Giorgetti, A. Bemporad, H. E. Tseng, and D. Hrovat, "Hybrid model predictive control application towards optimal semi-active suspension," *International Journal of Control,* vol. 79, no. 05, pp. 521-533, 2006.

[39] F. D .Torrisi and A. Bemporad, "HYSDEL-a tool for generating computational hybrid models for analysis and synthesis problems," *IEEE transactions on control systems technology,* vol. 12, no. 2, pp. 235-249, 2004.

زیرنویس‌ها

[1] Discrete Event Systems (DES)
[2] Piece-Wise Affine (PWA)
[3] Mixed Logical Dynamical (MLD)
[4] Max-Min-Plus-Scaling
[5] Model Predictive Control (MPC)
[6] Receding Horizon Control (RHC)
[7] Hybrid Model Predictive Control (H-MPC)
[8] Mixed Integer Program (MIP)
[9] Terminal Equality Constraint
[10] Equilibrium Pair
[11] Feasibility Problem
[12] Mixed Integer Quadratic Program (MIQP)
[13] Polynomial Time
[14] Branch and Bound (BnB)
[15] Quarter Car
[16] Tier Deflection
[17] Unsprung Mass Velocity
[18] Suspension Deflection
[19] Sprung Mass Velocity
[20] Ordinary H-MPC